\documentstyle[prd,aps,preprint]{revtex}
\begin{document}
\draft
\preprint{Nisho-3} \title{X Rays from Old Neutron Stars\\ Heated by
Axion Stars} 
\author{Aiichi Iwazaki}
\address{Department of Physics, Nishogakusha University, Chiba
  277-8585,\ Japan.} \date{Jun 3, 1999}  \maketitle
\begin{abstract}
We show that axionic boson stars collide with isolated 
old neutron stars with strong 
magnetic field ($>10^8$ Gauss)
and causes the neutron stars to radiate X ray by heating them.
Surface temperatures of 
such neutron stars becomes $10^5\,\mbox{K}
\sim 10^6\,\mbox{K}$.
We suggest that these are possible candidates for X ray sources 
observed in ROSAT Survey.
We discuss a possible way of identifying such neutron stars.
We also point out that the collision generates a burst of monochromatic 
radiations with frequency given by axion mass.
\end{abstract}
Axion, Neutron Star, Boson Star, Dark Matter, X Ray 
\hspace*{3cm}
\vskip2pc
Axions\cite{PQ} are one of most plausible candidates of 
dark matter in the Universe. 
The axions are produced \cite{kim,text} in early Universe 
mainly by decay of axion strings, decay of axion domain wall 
or coherent oscillation of axion field. 
These axions can form\cite{kolb} coherent axionic boson stars\cite{re} 
in the present Universe. Namely, some of the axions 
contract themselves gravitationally to axionic boson stars.
We call them simply as axion stars.  

In the previous papers\cite{iwa1,iwa2} 
we have pointed out that when axion stars collide
with cold white dwarfs invisible with present observational apparatus,
they heat such white dwarfs and make them visible.
This heating arises owing to the energy 
deposited by the axion
stars to the white dwarfs. That is, magnetic field of the white dwarfs
induces an electric field in the axion stars, which turns to generate   
an electric current in the white dwarfs. The energy of this electric current 
is dissipated, owing to finite electric conductivity 
of the white dwarfs.
Consequently the white 
dwarfs gain thermal energy, that is, the energy of the axion stars 
is transformed to the thermal energy of the white dwarfs by the collision. 
As a result, old white dwarfs regain their brightness. 
This thermalization of the axion energy under the magnetic field  
is a phenomenon similar to one arising 
in a cavity proposed for detection of the axion by Sikivie\cite{S}.
   
In this letter we point out that cold neutron stars ( NSs ) 
also regain brightness by collision with axion stars. 
As a result these neutron stars becomes detectable 
with X ray observation. In particular, isolated old neutron stars
become to emit X ray with this mechanism just as those accreting 
gas of interstellar medium. Hence 
they are possible candidates for X ray sources detected 
in ROSAT ALL-Sky Survey. 
Since both values of strength of magnetic field and electric conductivity
are extremely large in the case of NS, energy dissipation of axion star
proceeds very rapidly so that the collision would be observed as an 
explosion generating a blast of wind. Since electric currents induced  
in NS by the collision are oscillating with single frequency, 
a burst of monochromatic radiations is also produced.

We calculate 
luminosity of such NSs and estimate the number of them 
existing in the neighborhood of the sun.
We show that old NSs gain so much energies  
with the collision as for their surface temperatures to increase
up to $10^5\mbox{K}\sim 10^6$K.  
Furthermore, we show that  
there may be one or more such NSs within the distance of 
$1$Kpc around the sun, assuming dark matter 
being dominated by axion stars.
However, the precise number 
depends on several unknown parameters, e.g. mass of axion stars, 
collision parameters between axion star and NS, etc.
The masses of the axion stars in which we are interested are assumed such as
$M_a=10^{-11}M_{\odot}\sim 10^{-13}M_{\odot}$, which have been favored
according to arguments of the generation 
mechanism of axion stars by Kolb and Tkachev\cite{kolb}

Let us first explain briefly axionic boson stars\cite{iwa1} 
and then we explain\cite{iwa2} 
how they dissipate their energies in NS.
In general, boson stars are composed of coherent bosons 
bounded gravitationally, which 
are described by a solution of a corresponding boson field equation
coupled with gravity. 
In our case axions are
such bosons with mass $m_a$ and are represented by a real scalar field $a$.
Axion stars are coherent bound states of the boson and 
are characterized by their mass $M_a$ or radius $R_a$,
which are related with each other. Explicitly they are represented 
approximately by 

\begin{equation}
\label{a}
a=f_{PQ}a_0\sin(m_at)\exp(-r/R_a)\,, 
\end{equation}
where $t$ ($r$) is the time (radial) coordinate and 
$f_{PQ}$ is the decay constant of the axion. 
The value of $f_{PQ}$ is constrained from cosmological 
and astrophysical considerations\cite{text} such as 
$10^{10}$ GeV $< f_{PQ} <$ $10^{12}$ GeV. Here dimensionless amplitude, 
$a_0$ in eq(\ref{a}) is represented explicitely 
in terms of the radius, $R_a$
in the limit of small mass ( e.g. $\sim 10^{-12}M_{\odot}$)
of the axion star\cite{iwa1},

\begin{equation}
\label{a_0}
a_0=1.73\times 10^{-8} \frac{(10^8\,\mbox{cm})^2}{R_a^2}\,
\frac{10^{-5}\,\mbox{eV}}{m_a}\quad.
\end{equation}
In the same limit we 
have found\cite{iwa1} a simple relation between the mass $M_a$ 
and the radius $R_a$ 
of the axion star, 

\begin{equation}
\label{mass}
M_a=6.4\,\frac{m_{pl}^2}{m_a^2R_a}\,,
\end{equation} 
with Planck mass $m_{pl}$.
Numerically, 
$R_a=1.6\times10^8\,M_{12}^{-1}\,m_5^{-2}\,\mbox{cm}$ 
where $M_{12}\equiv M_a/10^{-12}\,M_{\odot}$
and $m_5\equiv m_a/10^{-5}\,\mbox{eV}$. A similar formula holds 
even without the limit but
with a minor modification of numerical coefficient.
It turns out that the axionic boson stars are ``oscillating'' 
with the frequency 
of $m_a/2\pi$\cite{real}. It has been shown that 
there are no physically relevant, ``static'', axionic boson stars. 
This property is specific in real scalar field. 
Static solutions of complex boson field representing boson stars 
are well known to exist\cite{re}.

We comment that there is a critical mass $M_c$ of axion star beyond which 
stable solutions do not exit. It is approximately given by 
$M_c\sim 10^{-5}M_{\odot}/m_5$. This notion is a similar to the critical mass 
of neutron stars or white dwarfs. The mass gives a typical mass scale of
these stars present in the Universe. 
Thus axion stars could have the mass of the order 
of such a critical mass, although in this paper we address axion stars
with the masses mentioned above. 

These axion stars induce electric fields under a magnetic field; the magnetic
field is supposed to be associated with neutron star in this paper. 
This can be 
easily seen by taking account of a following interaction term 
between axion and electromagnetic field,

\begin{equation}
   L_{a\gamma\gamma}=c\alpha a\vec{E}\cdot\vec{B}/f_{PQ}\pi
\label{EB}
\end{equation}
with $\alpha=1/137$, where 
$\vec{E}$ and $\vec{B}$ are electric and magnetic fields respectively. 
The value of $c$ depends on the axion models\cite{DFSZ,hadron};
typically it is of order 1.
It follows from this interaction that Gauss's law is modified such as

\begin{equation}
\label{Gauss}
\vec{\partial}\vec{E}=-c\alpha \vec{\partial}(a\vec{B})/f_{PQ}\pi
+\mbox{``matter''}
\end{equation}
where the last term, ``matter'', denotes contributions from ordinary matter.
The first term on the right hand side 
represents an electric charge density formed by axion field under 
external magnetic field $\vec{B}$\cite{Si}.
It is interesting that the axion field can induces the electric charge,
inspite of the field itself being neutral.  

We find that axion star induces an electric field,
$\vec{E_a}=-c\alpha a\vec{B}/f_{PQ}\pi$, under the magnetic field. 
This field is oscillating since the field $a$ is oscillating, 
and induces oscillating electric currents 
in NS. Accordingly, monochromatic radiations are emitted. 
Note that the radius $R_a$ of the axion star is much larger than the 
radius, $R_n\sim 10^6$cm, of neutron 
star; $R_a=10^7\,\mbox{cm}\sim 10^9\,\mbox{cm}$ for axion stars with mass,
$M_a=10^{-11}\,M_{\odot}\sim 10^{-13}\,M_{\odot}$. 
Hence the electric field induced at any place in 
the axion star does not necessarily generate electric current.
Electric current is only induced in electric conducting medium such as 
NS. Thus only a part of the axion star contacting NS
generates electric current in NS. 
This electric currents, $J_a=\sigma E_a$, are very strong 
since the electric conductivity, $\sigma$,
is quite high in NS ( for example, $\sigma \sim 10^{26}$/s 
in crystallized crust of NS\cite{con} ). Accordingly, the rate of 
the energy dissipation 
of the current is very large. Since the energy of the current is supplied by 
the axion star,    
the energy dissipation of the axion star itself proceeds very rapidly.  
Actually we find that axion stars dissipate their energies 
so rapidly that they evapolate quite soon simply when they touch 
with NS. It may be observed as an explosion generating 
a burst of monochromatic radiations as well as a blast of wind.
In this way, the axion star releases the entire energy ($\sim 10^{42}$ erg )  
in NS. The energy heats up cold NS.
Consequently such a NS becomes bright again.

Now we estimate the rate of the energy dissipation with use of 
Ohm's law. We consider the circumstance that axion star collides with NS,
which gains thermal energy inside of the axion star.  
Taking account of the fact that the radius $R_a$ 
of the axion star is much 
larger than that of NS, we calculate Joule's heat $W$ produced 
in NS,

\begin{eqnarray}
W&=&\int_{r<R_n}\sigma E_a^2\,dx^3\,,\\
&=&\sigma \alpha^2c^2B^2R_a^3a_0^2/8\pi\,,\\
&=&4c^2\times 10^{54}\,\mbox{erg/s}\,\frac{\sigma}{10^{26}/s}\,
\frac{M}{10^{-12}\,M_{\odot}}\,\frac{B^2}{(10^{12}\,\mbox{G})^2}\,,
\end{eqnarray}  
with $c\sim 1$, where we have used eqs(\ref{a})$\sim $(\ref{mass})
and have supposed that strength of magnetic field
of NS is typically given by $10^{12}$ Gauss.  
Here $\sigma$ is taken as an average conductivity in NS. Thus its value 
is smaller by $3$ order of magnitude than the value $10^{26}$/s quoted.
This is because electrons exist mainly in crust of NS whose size is given 
approximately by $R_n/10$.

This large rate of the energy dissipation implies 
rapid evapolation of the axion star when both two stars collide.
Actually, since the energy dissipation only arises in  
a part of the axion star contacting NS and 
the energy stored in the part is only about 
$M_a(R_n/R_a)^3\sim 10^{36}\,\mbox{erg}M_{12}^4m_5^6$, 
formally it takes $10^{36}/10^{54}=10^{-18}$ second 
for the dissipation of the energy ( it takes $10^{-6}$ second even with
$\sigma =10^{22}$/s and $B=10^8$ G ).
Therefore we find that 
the rapid evapolation of the axion energy arises even if 
NS possesses much weak magnetic field 
such as $10^8$G; the strength of this order of magnetic field 
is expected to be associated with old NSs.
Probably, such rapid dissipation of the energy may be seen as an explosion 
of envelope of NS\cite{iwa3}. 
It would generates a blast of wind, which subsequently collides with 
interstellar medium. Thus the medium emits radiations with various frequencies
as a burst.

We should mention that since electric currents induced in NS are oscillating
with frequency $m_a/2\pi$, the currents generate radiations 
with the frequency. Thus we expect that a burst of photons with energy $m_a$
of axion mass is produced by the collision.
This detection of the radiations is strong indication of 
the occurrence of such a collision. Furthermore, 
we can determine the mass 
of the axion by the detection of the burst.

After axion star collides with NS, it seems that axion 
star simply passes NS only with the loss of a part of its energy
through the dissipation mentioned above.
But it is reasonable to suppose
that it is trapped to NS, because the mass of the axion 
star is much smaller than that of the NS. Both the kinetic energy and 
the angular momentum of  
axion star would be dissipated through the above mechanism\cite{iwa2}. 
When the axion 
star is trapped, the entire energy of the axion star is dissipated after all.

Hereafter, 
we assume that the entire energy of the axion star is dissipated when 
it collides with NS. Hence, the energy gained by 
NS with the collision is given by the mass of the 
axion star, $M_a=10^{41}\,\mbox{erg}\sim 10^{43}\,\mbox{erg}$.
This energy heats up cold NSs 
and makes them become bright again.

In order to calculate roughly how temperature of such NSs rises up,
we use heat capacity of free nucleons composing the NSs for simplicity.
Assuming that the density of the NS is low enough for non-relativistic 
approximation to be valid, we may use the following formula 
of thermal energy\cite{text2} in the NS,

\begin{equation}
U=6\times 10^{47}\,\mbox{erg}\,(M/M_{\odot})\,(\rho/\rho_n)^{-2/3}\,
(T/10^9\,\mbox{K})^2,
\end{equation}
with $\rho_n=2.8\times 10^{14}\,\mbox{g cm}^{-3}$ being density of nucleon,
where $T$, $M$ and $\rho$ denote the core temperature, mass and 
average density of the NS, respectively.
Explicitly, we take the numerical parameters, $M=1.5\,M_{\odot}$ and 
$\rho=7\times 10^{14}\,\mbox{g cm}^{-3}$, which implies that the radius 
of the neutron star is given by $R=10^6$ cm. 

Since old NSs with their ages $\sim 10^{10}$ years have lost
almost all thermal energies, the energy deposited by the axion star 
is the main thermal energy by which their temperature is determined, 

\begin{equation}
T\simeq 8.6\times 10^6\mbox{K},\quad 2\times 10^6\mbox{K},\quad
\mbox{and}\quad 6\times 10^5\mbox{K}  
\end{equation}
for 
\begin{equation}
M_a=10^{-11}\,M_{\odot},\quad 10^{-12}\,M_{\odot},
\quad \mbox{and}\quad 10^{-13}\,M_{\odot}\quad \mbox{respectively}.
\label{M}
\end{equation}
 
This temperature is core temperature of NS.
In order to obtain luminosity of NS, 
we need to know surface temperature.
The temperature depends on not only the core temperature but also 
composition of atmosphere, or envelope of the NS.
Here we use a model\cite{model} 
in which it is assumed that the envelope is composed only of 
iron. Then we can find surface temperatures, $T_s$, 

\begin{equation}
T_s\simeq 2.8\times 10^5\mbox{K},\quad 1.4\times 10^5\mbox{K},
\quad \mbox{and}\quad 9\times 10^4\mbox{K}
\end{equation}
with which the luminosity
of NS is obtained,     

\begin{eqnarray}
\label{L}
L&=&7\times 10^{36}\,\mbox{erg/s}\,(T_s/10^7\mbox{K})^4\\
 &\simeq& 4.3\times 10^{30}\,\mbox{erg/s},\quad 2.7\times 10^{29}\,
\mbox{erg/s},
\quad \mbox{and}\quad 4.6\times 10^{28}\,\mbox{erg/s}
\end{eqnarray}
corresponding to the masses in eq(\ref{M}) of the axion stars, respectively.
From these luminosities we can estimate roughly a period of NSs keeping 
this brightness, that is the period of NSs exhausting the energies 
deposited by the axion stars. It is approximately 
given by $M_a/L\simeq 10^5$ years 
for any cases
mentioned above. This value is a lower limit of the 
period. Actual time scale for NS to exhaust the energy deposited
is longer than $10^5$ years. But we may think the value as a typical time 
scale during whose period NS maintains its brightness.

We have obtained the above values of the surface temperatures by assuming 
strength of surface gravity and composition of NS's envelope. 
The surface gravity 
is determined by the parameters we have used; 
$M=1.5M_{\odot}$ and $R=10^6$ cm. On the other hand, 
NS's envelope has been assumed to be
composed only of iron. If the surface gravity is much stronger or there are  
even few contamination of H or He in the envelope\cite{model}, 
the surface temperatures 
become much larger than the values estimated above.
Thus we may expect that real temperatures range roughly in
$10^5\,\mbox{K}\sim 10^6\,\mbox{K}$.
Therefore, these NSs may be possible candidates of 
X ray sources observed in ROSAT Survey.

Although the luminosities obtained are sufficiently large for observation,
it is hard to detect such NSs if the number of the NSs present in our 
neighborhood is quite few.  
Thus we wish to estimate how many such bright NSs are present in our 
galaxy. Especially we are concerned with the number of the NSs located within 
the distance of $1$ Kpc around the sun.  

In order to estimate the quantities, we need to know both numbers of cold
NSs and of axion stars present in our galaxy. 
The number of the NSs has been guessed on the basis
of present rate of appearance of supernova in a galaxy and abundance of 
heavy elements in our galaxy. The number has been estimated to be   
of the order of $10^9$ in our galaxy.
On the other hand the number of axion stars is completely unknown.
However, we may assume that 
the halo is composed mainly of the axion stars, because the axions 
are plausible candidates of the dark matter.
Using these assumptions, we can estimate the number of the NSs which 
have collided with axion stars and have not yet lost their brightness.

Since local density $\rho_a$ of halo is given approximately such that 
$\rho_a=0.5\times10^{-24}$g $\mbox{cm}^{-3}$\cite{text}, 
we find that the number 
density $n_a$ of axion stars is, 
$n_a=\rho \times (1\mbox{Kpc})^3/M_a\sim 6\times 10^{18}/M_{12}$ per 
$1\mbox{Kpc}^3$. Under the assumptions that cross section of 
the collision between a NS and an axion star is naively given by 
the geometrical cross section of axion star, $\pi R_a^2$ and that  
velocity of axion stars in halo is given typically by
$3\times 10^7$ cm/s, we calculate the number of the collisions 
per $1\mbox{Kpc}^3$ and per year,

\begin{equation}
R_c=n_a\times n_{ns}\times \pi R_a^2 v\times 1\,\mbox{year}
\simeq 10^{-8}\times \frac{1}{M_{12}^3 m_5^4}\,
\,\, \mbox{per year and per $1\mbox{Kpc}^3$}
\end{equation}
where $n_{ns}$ denotes the number of NSs in the volume of $1\mbox{Kpc}^3$. 
Here we have assumed uniform distribution of $10^9$ NSs 
in the disk of our galaxy.
This $R_c$ represents the production rate of NSs heated by the collision.
On the other hand, life time for such NSs to maintain 
brightness eq(\ref{L}) 
is about $10^5$ years. Thus, 
the number of the bright NSs produced for $10^5$ years 
by the collision is 
$\simeq 10^{-3}/M_{12}^3m_5^4$
in a local region with its volume $1\mbox{Kpc}^3$ around the sun.
In other words there are $\sim 1 /M_{12}^3m_5^4$ NSs in our galaxy.
The result suggests that if the mass of the axion star is smaller than
$10^{-13}M_{\odot}$, the number of NSs within the volume is larger than $1$ 
and hence such NSs may be detectable. Here we have restricted  
that $m_a >10^{-5}$ eV.

In the estimation of $R_c$ we have assumed naive geometrical cross 
section of the axion star as the cross section of the collision with NS. 
But since NS and axion star interacts gravitationally with each other, 
its collision cross section is larger than the naive one. 
Since the rate $R_c$ becomes larger as the cross section increase more, 
it is plausible that real number of the 
NSs is larger than the value obtained above.

Therefore we conclude that if the mass of the axion star colliding NS 
is smaller than $10^{-13}M_{\odot}$, 
or collision cross section is much larger than
the geometrical one $\sim 10^8\,M_{12}^{-1}\,m_5^2$ cm of the axion star,
the number of the NSs present within the distance of $1$Kpc around the sun
is large enough for observation. Since surface temperatures of these NSs
are of the order $10^5\mbox{K}\sim 10^6\mbox{K}$,
they are observable as X ray sources. 
In the derivation of this conclusion, 
significant assumption is that halo is composed mainly of axion stars. 
On this point, recent gravitational microlense observation\cite{macho} 
indicates that
a half of the halo is composed of objects with mass 
$0.1\,M_{\odot}\sim 0.5\,M_{\odot}$. If this is true, in order to obtain 
the conclusion we need 
the assumption that the other half of the 
halo is composed of axion stars.

Here we should comment that if masses of axion stars present 
in the Universe are of the order 
of the critical mass $M_c$, the energy released by the collision amounts
to $\sim 10^{49}\mbox{erg}/m_5$. Thus if axion mass $m_a$ can take 
a much smaller value such as $10^{-9}$eV, the energy reaches one observed
in gamma ray burst. Actually axion originated 
in string theory\cite{string} may take 
such a small value as its mass without contradicting observation. Then 
it is reasonable to guess\cite{iwa3} 
that such a collision between NS and axion star with $M_a\simeq M_c$
is a possible mechanism generating the gamma ray burst. 
The collision generates a burst of monochromatic radiations.  
Therefore, we expect to detect the burst of such radiations associated with
gamma ray burst. Wave length of the radiations is 
of the order $10^4\mbox{cm}\sim 10^5\mbox{cm}$.

Finally we discuss how we should identify NSs as ones heated by axion stars.
Especially, we wish to distinguish them from NSs which emit X ray by 
accreting gas of interstellar medium. The latter is located in a region 
where density of the gas is relatively large, while the former is located 
even in a region without any gas. Furthermore, the NS accreting the gas must
have very low velocity ($\sim 10\mbox{km/s}$), 
while the NS heated by axion star may have relatively 
high velocity for instance $\sim 200\mbox{km/s}$. 
Accordingly, if a X ray source is located in a region with few interstellar
medium and has a relatively high velocity, it may be identified as a 
NS heated by axion star. However, such a NS may be a middle aged NS which also
can emit X rays. For further distinction, we note that
the collision between 
NS and axion star would be seen as an explosion  
generating a blast of wind as well as a burst of monochromatic radiations.
Thus the NS having collided with axion star 
would have a cloud of gas surrounding it, which was carried by the blast 
of the wind. This kind of the cloud is not present around 
middle aged NSs as well as NSs accreting the gas of the interstellar medium. 
Hence if we detect such a cloud
around a X ray source, the source is 
strong candidate of the NS heated by axion star.

The author wish to express his thank for the hospitality in Tanashi KEK.





\end{document}